\documentclass[sigconf]{acmart}

\usepackage{algorithm}
\usepackage{algorithmic}
\def\BibTeX{{\rm B\kern-.05em{\sc i\kern-.025em b}\kern-.08emT\kern-.1667em\lower.7ex\hbox{E}\kern-.125emX}}
    
\copyrightyear{2019} 
\acmYear{2019} 
\setcopyright{acmcopyright}
\acmConference[KDD '19]{The 25th ACM SIGKDD Conference on Knowledge Discovery and Data Mining}{August 4--8, 2019}{Anchorage, AK, USA}
\acmBooktitle{The 25th ACM SIGKDD Conference on Knowledge Discovery and Data Mining (KDD '19), August 4--8, 2019, Anchorage, AK, USA}
\acmPrice{15.00}
\acmDOI{10.1145/3292500.3330651}
\acmISBN{978-1-4503-6201-6/19/08}

\begin{document}

%
\title[MOBIUS: Next Generation of Query-Ad Matching in Baidu's Sponsored Search]{MOBIUS: Towards the Next Generation of Query-Ad Matching in Baidu's Sponsored Search}

%

\author{$^1$Miao Fan, $^2$Jiacheng Guo, $^2$Shuai Zhu,  $^2$Shuo Miao, $^1$Mingming Sun, $^1$Ping Li}

\email{{fanmiao, guojiacheng, zhushuai, miaoshuo, sunmingming01, liping11}@baidu.com}

\affiliation{
  \institution{$^1$ Cognitive Computing Lab (CCL), Baidu Research, Baidu Inc.}
  \institution{$^2$ Baidu Search Ads (Phoenix Nest), Baidu Inc.}
}

%
\renewcommand{\shortauthors}{M. Fan, J. Guo, S. Zhu, S. Miao, M. Sun, and P. Li}

%
\begin{abstract}
Baidu runs the largest commercial web search engine in China, serving hundreds of millions of online users every day in response to a great variety of queries. In order to build a high-efficiency sponsored search engine, we used to adopt a three-layer funnel-shaped structure to screen and sort hundreds of ads from billions of ad candidates subject to the requirement of low response latency and the restraints of computing resources. Given a user query, the top matching layer is responsible for providing semantically relevant ad candidates to the next layer, while the ranking layer at the bottom concerns more about business indicators (e.g., CPM, ROI, etc.) of those ads. The clear separation between the matching and ranking objectives results in a lower commercial return. The Mobius project has been established to address this serious issue. It is our first attempt to train the matching layer to consider CPM as an additional optimization objective besides the query-ad relevance, via directly predicting CTR (click-through rate) from billions of query-ad pairs. Specifically, this paper will elaborate on how we adopt active learning to overcome the insufficiency of click history at the matching layer when training our neural click networks offline, and how we use the SOTA ANN search technique for retrieving ads more efficiently (Here ``ANN'' stands for approximate nearest neighbor search).  We contribute the solutions to Mobius-V1 as the first version of our next generation query-ad matching system.
\end{abstract}


\keywords{Sponsored search; query-ad matching; active learning; click-through rate (CTR) prediction; approximate nearest neighbor (ANN) search}

\begin{teaserfigure}
\begin{center}
  \includegraphics[width=0.75\textwidth]{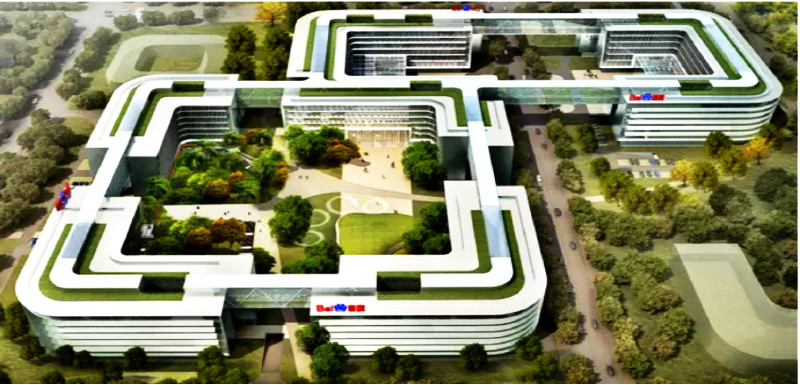}
 \end{center}
  \vspace{-0.15in}
  \caption{``Mobius'' is the internal code name of this project. Coincidentally, the well-known  ``Mobius Loop'' is also the bird's-eye view of Baidu's Technology Park in Beijing, China. }
  \label{fig:baidu}\vspace{0.1in}
\end{teaserfigure}

\maketitle
\section{Introduction}
\begin{figure}[h!]
    \centering
  \includegraphics[width=0.38\textwidth]{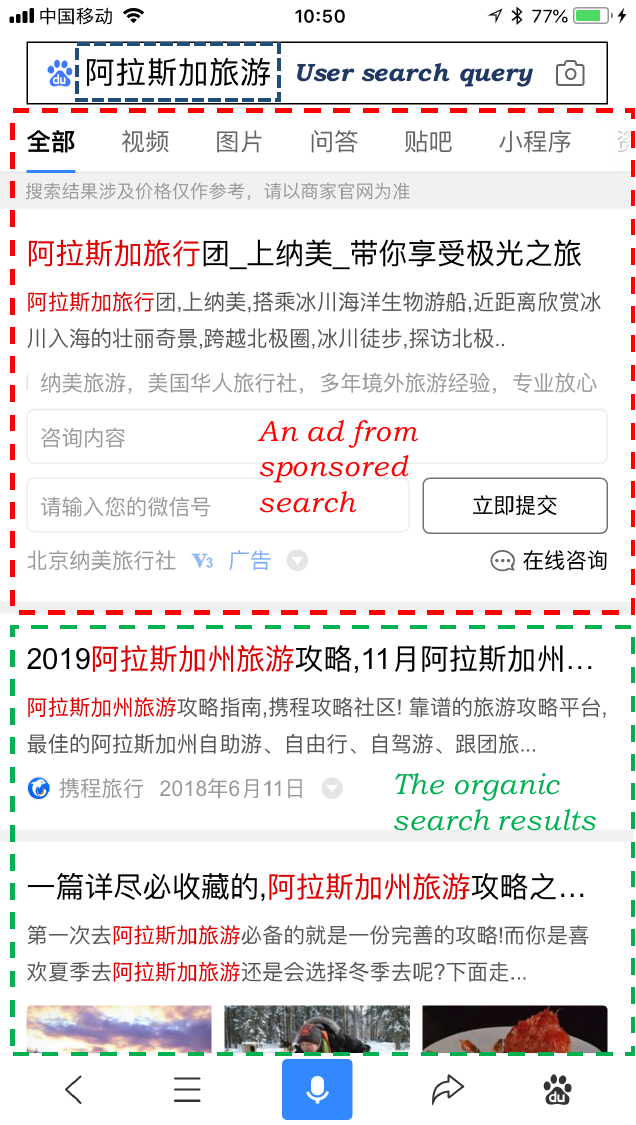}
  \vspace{-3mm}
  \caption{A screen-shot of Baidu Search results on mobile phones given a search query ``tourism in Alaska'' (in Chinese). Our sponsored search engine is in charge of providing helpful ads on each page before the organic search results.}
  \label{fig:baidu_sponsored_search}
\end{figure}

Baidu Search (\url{www.baidu.com}), as the largest commercial search engine in China, daily serves hundreds of millions of online users in response to a great variety of search queries. It is common knowledge that advertising has been the main revenue source for all major commercial search engine firms in the world. In this paper, we focus on explaining some of the recent exciting development and invention in Baidu's Search Ads system (conventionally known as the ``Phoenix Nest'' inside Baidu).  As shown by Figure~\ref{fig:baidu_sponsored_search}, it plays a vital role in retrieving advertisements (ads) which are relevant to user queries to attract clicks as advertisers are willing to pay when their ads get clicked. The goal of Baidu sponsored search system is to form and nourish a virtuous circle among online users, advertisers, and our sponsored search platform. 

\begin{figure}[h!]
    \centering
  \includegraphics[width=0.45\textwidth]{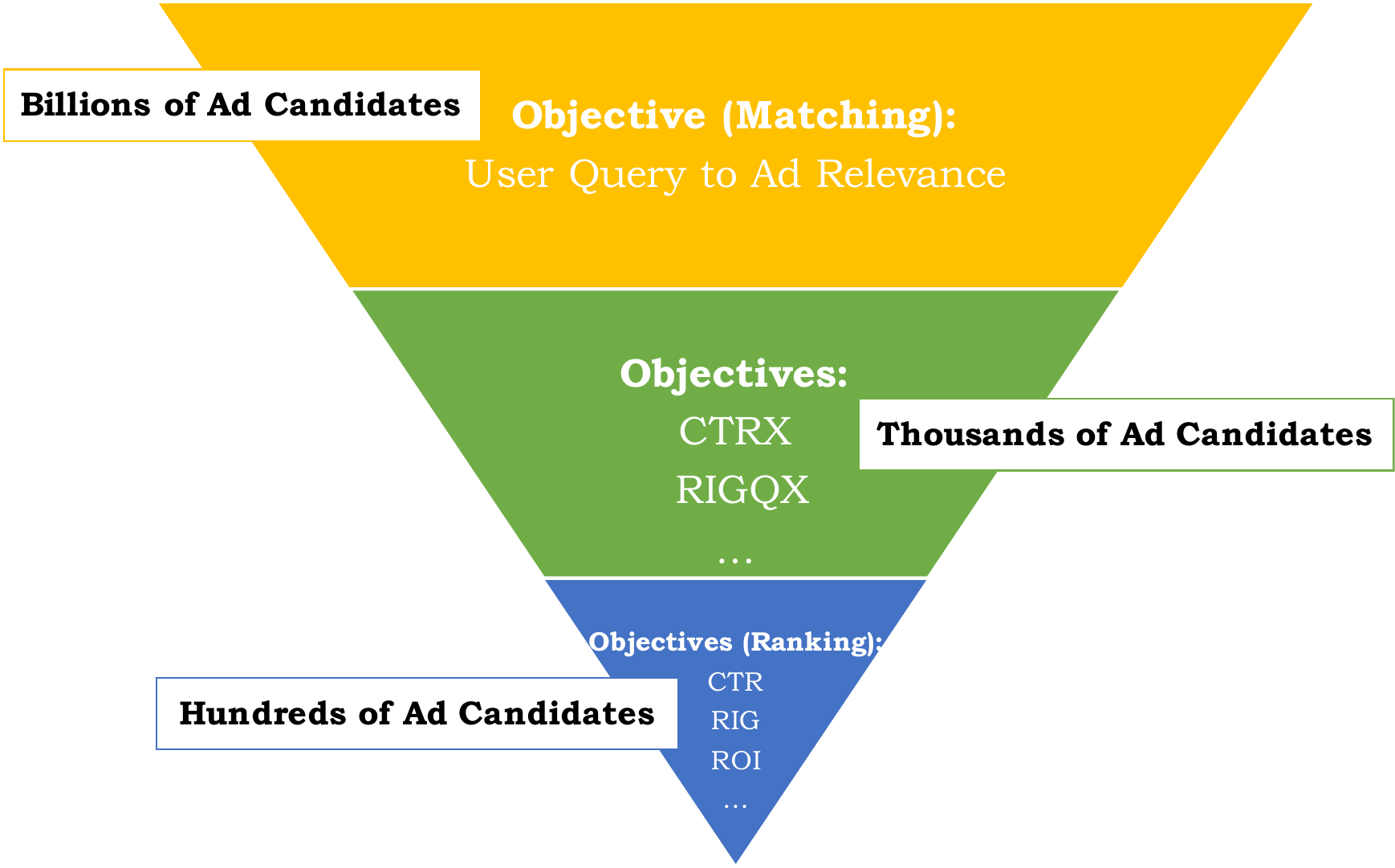}
    \vspace{-3mm}
  \caption{The three-layer funnel-shaped structure of the previous sponsored search system in Baidu. Given a user query, it is highly efficient to retrieve hundreds of relevant and high-CPM ads from billions of ad candidates.}
  \label{fig:funnel}
  \vspace{-0.1in}
\end{figure}

\begin{figure*}
    \centering
  \includegraphics[width=0.9\textwidth]{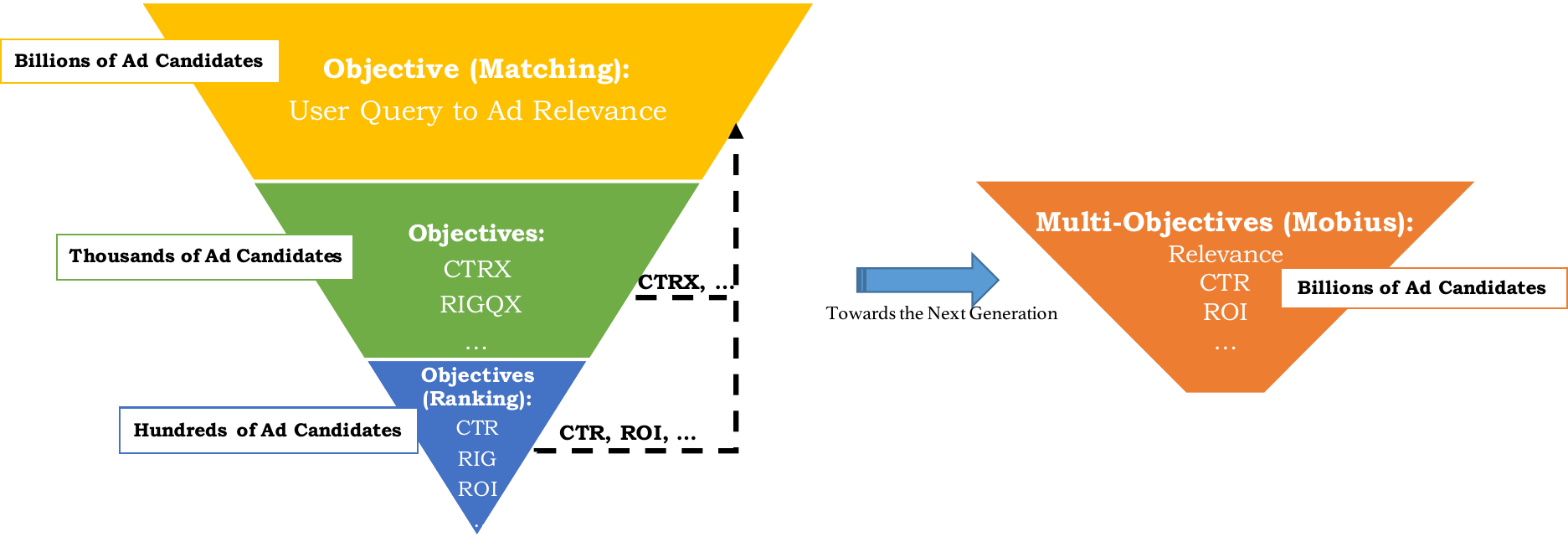}
  \vspace{-3mm}
  \caption{The distinct objectives of matching and ranking layer lead to lower CPM which is one of the key business indicators of a sponsored search engine. Therefore, we are engaged in building a high-efficient query-ad matching system (i.e., Mobius) in Baidu sponsored search. Mobius is expected to unify the learning objectives of the query-ad relevance and many other business indicators together, subject to lower response latency, limitation of computation resources and tiny adverse impact on user experience. For now, we have deployed the first version of Mobius (Mobius-V1) which can more accurately predict CTR for billions of user query \& ad pairs.}
  \vspace{-0.1in}
  \label{fig:mobius}
\end{figure*}

Conventional sponsored search engines \cite{fain2006sponsored,jansen2008sponsored,Dave:2014:CAT:2692913.2692914} generally display ads through a two-step process. The first step is to retrieve relevant ads given a query, and the next step is to rank these ads based on predicted user engagement. As a high-efficiency sponsored search engine for commercial use in Baidu, we used to adopt a three-layer funnel-shaped structure to screen and sort hundreds of ads from billions of ad candidates subject to the requirement of low response latency and the restraints of computing resources. As illustrated in Figure~\ref{fig:funnel}, the top matching layer is responsible for providing relevant ad candidates to the next layer given a user query and the rich profile of the user. To cover more semantically relevant ads, query expansion \cite{abhishek2007keyword,broder2009online,DBLP:conf/kdd/BaiOZFRST18} and natural language processing (NLP) techniques \cite{baeza2008towards} are mostly leveraged. The ranking layer at the bottom concerns more about business indicators \cite{graepel2010web}, such as cost per mile (CPM $=$ CTR $\times$ Bid), return on investment (ROI), etc., of the filtered ads provided by the upper layer. 

However, this separation/distinction between matching and ranking objectives lead to a lower commercial return for various reasons. Given a user query, we have to employ complex models and to spend a lot of computing resources on ranking hundreds or even thousands of ad candidates. Perhaps most disappointingly, the ranking models report that many relevant ads are not offered by high CPM and will not be displayed.
To address this issue, Baidu Search Ads has set up the ``Mobius'' project which aims towards the next generation query-ad matching system in Baidu's sponsored search. This project is expected to unify the diverse learning objectives including the query-ad relevance and many other business indicators together, subject to lower response latency, restraints of computing resources and tiny adverse impact on user experience.

In this paper, we introduce Mobius-V1 which is our first attempt for teaching the matching layer to take CPM as an additional optimization objective besides the query-ad relevance. In other words, Mobius-V1 has the capability of accurately and rapidly predicting click-through rate (CTR) for billions of user query \& ad pairs. To achieve this goal, we must resolve the following major problems:
\begin{itemize}
    \item \textbf{Insufficient click history}: The original neural click model employed by the ranking layer was trained by high-frequency ads and user queries. It tends to estimate a query-ad pair at a much higher CTR for display once either a high-frequency ad or a high-frequency query appears, even though they might have low relevance.  
    \item \textbf{High computational/storage cost}: Mobius is expected to forecast multiple indicators (including relevance, CTR, ROI, etc.) of billions of user query \& ad pairs. It naturally faces the challenge of greater consumption on computing resources.
\end{itemize}

To address the problems above, we first design a ``teacher-student'' framework inspired by the idea of active learning \cite{wang2011active,settles2012active} to augment the training data for our large-scale neural click model to predict CTR for billions of user query \& ad pairs. Specifically, an offline data generator is responsible for constructing synthetic query-ad pairs given billions of user queries and ad candidates. These query-ad pairs are constantly judged by a teacher agent which is derived from the original matching layer and is good at measuring the semantic relevance of a query-ad pair. It can help detect the bad cases (i.e., high CTR but low relevance) in the synthetic query-ad pairs. Our neural click model, as a student, is taught by the additional bad cases to improve the ability of generalization on tail queries and ads. To save the computing resources and satisfy the requirement of low response latency, we further adopt the most recent state-of-the-art approximate nearest neighbor (ANN) search and Maximum Inner Product Search (MIPS) techniques for indexing and retrieving a large number of ads more efficiently. 

To tackle the aforementioned challenges, Mobius-V1, as the first version of our next generation query-ad matching system, is an integration of the solutions above and has already been deployed in Baidu's sponsored search engine. 


\section{Vision of Baidu's Sponsored Search}
For a long period of time, the funnel-shaped structure is a classical architecture of sponsored search engines \cite{fain2006sponsored,jansen2008sponsored,Dave:2014:CAT:2692913.2692914}. The major components include the query-ad matching and ad ranking. The query-ad matching is typically a lightweight module which measures the semantic relevance between a user query and billions of ads. In contrast, the ad ranking module should concern much more business indicators such as CPM, ROI, etc., and use complex neural models to sort hundreds of ad candidates for display. This decoupled structure is a wise option to save the expensive computing resources in the early days.
Moreover, it can also facilitate both scientific research and software engineering as the two modules can be assigned to different research/development teams to maximize individual objectives.  

Baidu's sponsored search used to adopt a three-layer funnel-shaped structure which is shown by Figure~\ref{fig:funnel}. The optimization objective of the top matching layer (denoted by $\mathcal{O}_{Matching}$) is to maximize the average relevance score among all the query-ad pairs:
\begin{equation}
\vspace{-2mm}
      \mathcal{O}_{Matching} = \max {\frac{1}{n} {\sum_{i=1}^n{\text{Relevance}~(query_i, ad_i)}}}.
\end{equation}
 
However, according to our long-term analysis on the performance of Baidu's sponsored search engine, we find out that the distinction/separation between matching and ranking objectives tends to lead to lower CPM which is one of the key business indicators for a commercial search engine. It is unsatisfactory when the models in the ranking layer report that many relevant ads provided by the matching layer will not be displayed on search results as they are estimated not to have higher CPM.

With the rapid growth of computing resources, the Baidu Search ads team (``Phoenix Nest'') has recently established the Mobius project which aims towards the next generation query-ad matching system in Baidu's sponsored search. The blueprint of this project as illustrated in Figure~\ref{fig:mobius} looks forward to unifying multiple learning objectives including the query-ad relevance and many other business indicators into a single module in Baidu's sponsored search, subject to lower response latency, limited computing resources and tiny adverse impact on user experience. 

This paper will report the first version of Mobius, i.e., Mobius-V1, which is our first attempt to teach the matching layer considering CPM as an additional optimization objective besides the query-ad relevance. Here we formulate the objective of Mobius-V1 as follows,
\begin{equation}
\begin{aligned}
      & \mathcal{O}_{Mobius-V1} = \max {\sum_{i=1}^n{\text{CTR}~(user_i, query_i, ad_i) \times Bid_i}}, \\
      & s.t. ~~\frac{1}{n} {\sum_{i=1}^n{\text{Relevance}~(query_i, ad_i) \geq threshold}}. \\
\end{aligned}
\label{eq2}
\end{equation}
Thus, it  becomes a challenge about how to accurately predict CTR for billions pairs of user quires and ad candidates in Mobius-V1. In the rest of the paper, we will describe how we design, implement, and deploy Mobius-V1, in great details. 

\section{MOBIUS: Next Generation Query-Ad Matching System}

``Mobius'' is Baidu's internal code name of this project. Coincidentally, the well-known ``Mobius Loop'' is also the bird's eye view of Baidu's Technology Park in Beijing, China; see Figure~\ref{fig:baidu}.   ``Mobius-V1'' is our first attempt (which has been successfully deployed) to transfer our neural click model to the matching layer directly facing billions of user query and ads.  As the scale of input data dramatically increases, we need to re-train our neural click model offline and update the techniques of indexing and retrieving ads.

\subsection{Active-Learned CTR Model}\label{sec:3.1}

\begin{figure}[h!]
    \centering
  \includegraphics[width=0.45\textwidth]{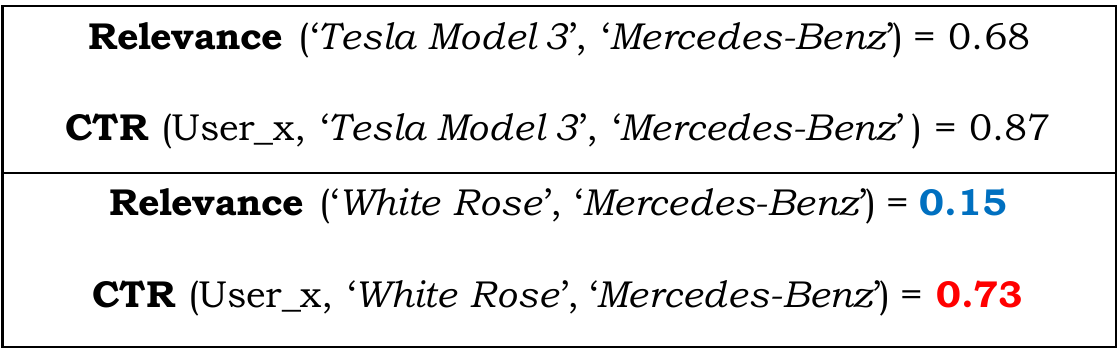}
    \vspace{-0.1in}
  \caption{An example of a bad case that the original CTR model could not handle well. As the neural click network employed by the ranking layer was originally trained by high-frequency ads and queries, it tends to estimate a query-ad pair at a higher CTR once a high-frequency ad (e.g., ``Mercedes-Benz'' in this case) appears, even though ``White Rose'' and ``Mercedes-Benz'' have little relevance.}
  \label{fig:ex}
\end{figure}

For over 6 years, Baidu's sponsored search engine has been using the deep neural networks (DNN) for the CTR model (of gigantic size). Recently, Mobius-V1 has adopted an innovative new architecture. An intuitive and simple way of building Mobius-V1 is to reuse the original CTR model in the ranking layer. It is a large-scale and sparse deep neural network (DNN) which is in favor of memorization. However, it suffers from a severe bias on CTR prediction of either the user queries or the ads in the tail. Consider, as shown in  Figure~\ref{fig:ex}, the two queries  ``Tesla Model 3'' and ``White Rose'' requested by the same user as in the search log. For the funnel-shaped structure adopted in the past, the relevance between the query ``Tesla Model 3'' and the ad ``Mercedes-Benz'' is firstly guaranteed by the matching layer. Then our neural click model in the ranking layer tends to predict a higher CTR on the query-ad pair as ``Tesla Model 3'' is a high-frequency query and leaves a rich click history on the ad ``Mercedes-Benz'' in our search log. However, in Mobius-V1 we attempt to use our neural click network to directly handle billions of query-ad pairs lacking the guarantee of relevance. It is natural that many irrelevant query-ad pairs come out (e.g., the query ``White Rose'' and the ad ``Mercedes-Benz'' in Figure~\ref{fig:ex}). Nevertheless, we have found out that our neural click model still tends to predict higher CTR for those irrelevant query-ad pairs.

According to our analysis on the query log in Baidu's sponsored search, the ads and the user queries suffer from the long-tail effect and the cold start problem. 
Therefore, we can not directly leverage the original neural click model to accurately predict CTR for billions of user queries and ads in the tail. The key to the problem is how we teach our model learning to recognize the ``low relevance but high CTR'' query-ad pairs as the bad cases. 

\begin{algorithm}[!h]
\caption{The active learning procedure for training neural click model to predict CTR for billions of query-ad pairs.}
\label{alg}
\begin{algorithmic} 
\REQUIRE Click\_History, Relevance\_Judger (Teacher), and Neural\_Click\_Model (Student)
\STATE \vspace{-0.07in}
\WHILE{$epoch \leq N$ or $err \geq \epsilon$}
\STATE\vspace{-0.07in}
\STATE \# Loading a batch of \textbf{Click\_History}
\STATE $data$=$\{(user_i, query_i, ad_i, (un)click_i), i = 1, 2, ..., n\}$ 

\STATE \vspace{-0.07in}
\STATE \# Building the query set and the ad set, respectively
\STATE $querySet$=Set(List($query_i$))
\STATE $adSet$=Set(List($ad_i$))

\STATE \vspace{-0.07in}
\STATE \# Generating the augmented data
\STATE $augData$=$querySet \otimes adSet$ 

\STATE \vspace{-0.07in}
\STATE \# Obtaining the low-relevance augmented data
\STATE $lowRelAugData$=\textbf{Relevance\_Judger}($augData$)

\STATE \vspace{-0.07in}
\STATE \# Obtaining the predicted CTRs for the low-relevance augmented data
\STATE $(lowRelAugData, pCtrs)$=\textbf{Neural\_Click\_Model}($lowRelAugData$)

\STATE \vspace{-0.07in}
\STATE \# Sampling the bad cases from low-relevance augmented data according the predicted CTRs
\STATE $badCases$=Sampling($lowRelAugData$) s.t. $pCTRs$

\STATE \vspace{-0.07in}
\STATE \# Adding the bad cases into the training buffer with the \textbf{Click\_History}
\STATE $trainBuffer$=$[data, badCases]$

\STATE \vspace{-0.07in}
\STATE \# Updating the weights inside \textbf{Neural\_Click\_Model} with the data in the training buffer
\STATE Updating (\textbf{Neural\_Click\_Model}) s.t. $trainBuffer$
\STATE\vspace{-0.07in}
\ENDWHILE
\end{algorithmic}
\end{algorithm}
\begin{figure*}
    \centering
  \includegraphics[width=0.92\textwidth]{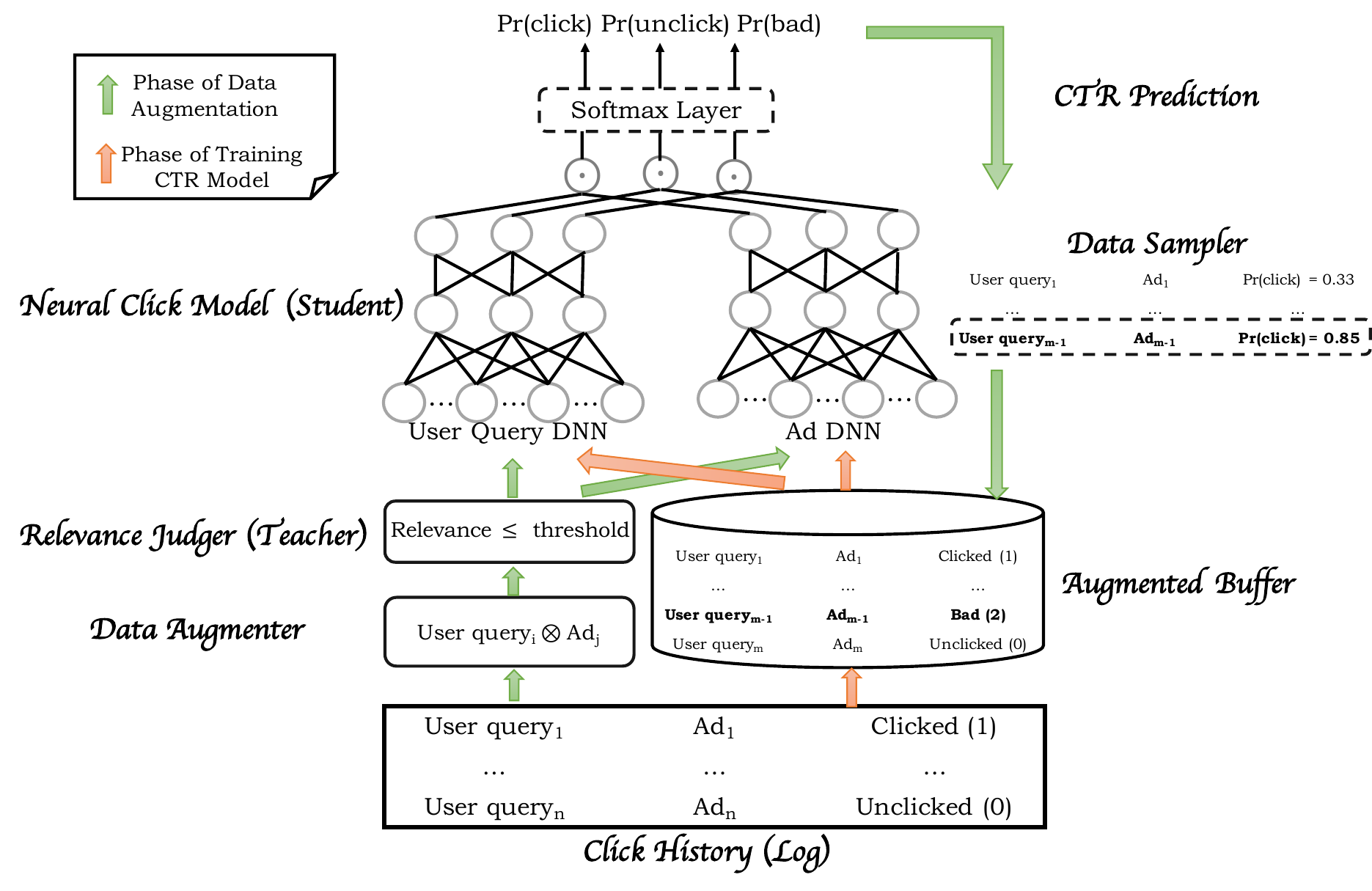}
  \vspace{-0.1in}
  \caption{The flow diagram of actively training our neural click model with augmented data. The iterative procedure has two phases: data augmentation and CTR model learning. The phase of data augmentation starts from loading a batch of click history (i.e., user query \& ad pairs) into a data augmenter. The data augmenter adopts a cross join operation to generate more user query \& ad pairs even though they do not appear in the click history. Then we bring in the original matching model as a teacher to grade the relevance of those pairs. The teacher set a threshold to retain the irrelevant query-ad pairs which are further fed into our neural click network. Our neural click network acts as a student and tries to predict the CTRs of query-ad pairs. A data sampler is responsible for sampling and labeling the bad cases (i.e., user query \& ad pairs with lower relevance but higher CTR). After the training buffer is augmented by the bad cases, we start the second phase to train our neural click model which predicts the probability of distribution in three categories: click, unclick and bad. Once the augmented data in the buffer have been used, we clean the buffer and wait for loading the next batch of click history.}
  \label{fig:ctr}
\end{figure*}

To solve this problem, we propose to use the original relevance judger in the matching layer as the teacher to make our neural click model aware of the ``low relevance'' query-ad pairs.
Our neural click model, as the student, acquires the additional knowledge on relevance from the augmented bad cases in an active learning fashion.
Figure~\ref{fig:ctr} illustrates the fashion by a flow diagram and Algorithm~\ref{alg} shows the training procedure of teaching our neural click model with active learning in pseudo code. 
Generally speaking, the iterative procedure of active learning has two phases: data augmentation and CTR model learning. To be specific, we will elaborate the modules in each phase step by step.

The phase of data augmentation starts from loading a batch of click history (i.e., user query \& ad pairs) from query logs into a data augmenter. Every time the data augmenter receives the query-ad pairs, it splits them into two sets: a query set and an ad set. Then we apply a cross join operation ($\otimes$) to the two sets for the sake of constructing more user query \& ad pairs. Suppose that there are $m$ queries and $n$ ads in the batch of click history, and then the data augmenter can help generate $m \times n$ synthetic query-ad pairs. After listing all possible query-ad pairs, the relevance judger involves in and takes charge of grading the relevance of these pairs. As we want to discover the low relevance query-ad pairs, a threshold is set to reserve those pairs as candidate teaching materials. These low relevance query-ad pairs, as teaching materials, are fed into our neural click model for the first time, and each pair is assigned with CTR predicted by the updated model in the previous iteration. To teach our 3-classes (i.e., click, unclick and bad) neural click model learning to recognize ``low relevance but high CTR'' query-ad pairs, we may intuitively set another threshold to filter out most low CTR query-ad pairs. However, we consider a better option to balance the exploration and exploitation of the augmented data. We employ a data sampler which selects and labels the augmented data referred to the predicted CTRs of those synthetic query-ad pairs. Once a query-ad pair is sampled as a bad case for our neural click network, this pair is labeled by an additional category, i.e., \textit{bad}. 

In the phase of learning our CTR model, both the click/unclick history and the labeled bad cases are added into the augmented buffer as the training data. Our neural click network is a large-scale and multi-layer sparse DNN which is composed of two subnets, i.e., user query DNN and ad DNN. As illustrated by Figure~\ref{fig:ctr}, the user query DNN on the left takes rich user profiles and queries as inputs and the ad DNN on the right regards the ad embeddings as features. Both subnets produce a distributed representation with $96$ dimensions each of which is segmented into three vectors ($32 \times 3$). We apply the inner product operation $3$ times to the three pairs of vectors between the user query DNN and ad DNN and adopt a softmax layer for CTR prediction.  

Overall, we contribute a type of learning paradigm to train our neural click model offline in Baidu's sponsored search engine. 
For the sake of improving its capability of generalization on CTR prediction for billions of query-ad pairs in the tail, the neural click model (student) can actively query the relevance model (teacher) for labels. This type of iterative supervised learning is known as active learning~\cite{wang2011active,settles2012active}.

\subsection{Fast Ads Retrieval}
\begin{figure}[h!]
    \centering
  \includegraphics[width=0.32\textwidth]{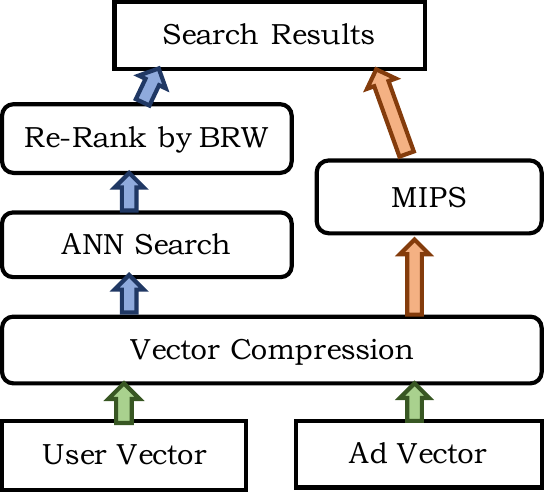}
  \caption{The fast ad retrieval framework. The two types of vectors will be compressed first to save the memory space. After that, two strategies can be applied: (a) ANN search by cosine similarity and then re-rank by the business related weight (BRW); (b) Ranking by exploiting weight, which is a Maximum Inner Product Search (MIPS) problem.}  \label{fig:ann}\vspace{-0.1in}
\end{figure}

In Baidu's sponsored search engine, we have been using the deep neural networks (i.e., user query DNN and ad DNN) illustrated by Figure~\ref{fig:ctr} to acquire both the embeddings of queries and ads, respectively. Given a query embedding, Mobius must retrieve the most relevant and the highest CPM ads from billions of ad candidates as stated in Eq.~(\ref{eq2}). Of course, it is unpractical to calculate it exhaustively for each query although the brute-force search can theoretically discover all the ads (i.e., 100\% ad recall) we are looking for. 
The online services often have restricted latency constraints and the ad retrieval must be done in a short period. Thus, we exploit approximate nearest neighbor (ANN) search techniques to speed up the retrieval process, as shown by Figure~\ref{fig:ann}. 

\subsubsection{ANN Search}
As shown in Figure~\ref{fig:ctr}, the mapping function combining user vectors and ad vectors by cosine similarity and then the cosine values go through a softmax layer to produce the final CTR. In this way, the cosine value and the CTR are monotonically correlated. After the model is learned, it will be clear they are positively or negatively correlated. If it is negatively correlated, we can easily transfer it to positively correlated by negating the ad vector. In this way, we reduce the CTR ranking problem into a cosine ranking problem, which is a typical ANN search setting.

The goal of approximate nearest neighbor (ANN) search is to retrieve, for a given query object, the ``most similar'' set of objects from a large corpus, by only scanning a small fraction of objects in the corpus. This is a fundamental problem and has been actively studied since the early days in computer science~\cite{Article:Friedman_75,Article:Friedman_77}. Typically, popular algorithms for ANN have been based on the idea of space-partitioning, including tree-based methods~\cite{Article:Friedman_75,Article:Friedman_77}, random hashing methods~\cite{DBLP:journals/cn/BroderGMZ97,Proc:Indyk_STOC98,gionis1999similarity, charikar2002similarity,Proc:Li_Owen_Zhang_NIPS12,Proc:Shrivastava_AISTATS14}, quantization based approaches~\cite{jegou2011product, ge2013optimized, wu2017multiscale}, random partition tree methods~\cite{Proc:Dasgupta_STOC08,DBLP:journals/algorithmica/DasguptaS15}, etc. For this particular problem (which deals with dense and relatively short vectors), we have found that random partition tree methods are fairly effective. There is a known implementation of random partition tree methods called ``ANNOY'', among other variants~\cite{Proc:Dasgupta_STOC08,DBLP:journals/algorithmica/DasguptaS15}.


\subsubsection{Maximum Inner Product Search (MIPS)}
In the above solution, the business-related weight information is considered after the user vector and ad vector matching. In practice, this weight is vital in ads ranking. To take into account the weight information earlier in the ranking, we formalize the fast ranking process by a weighted cosine problem as follows:
\begin{align} \label{eq:weightedcos}
     \cos(x, y)\times w
    = \frac{x^\top y\times w}{\|x\|\|y\|}=\left(\frac{x}{\|x\|}\right)^\top\left(\frac{y\times w}{\|y\|}\right),
\end{align}
where $w$ is the business related weight, $x$ is user-query embedding and $y$ is the ad vector. Note that the weighted cosine poses an inner product searching problem, often referred as Maximum Inner Product Search (MIPS)~\cite{shrivastava2014asymmetric}. In this line of work, multiple frameworks can be applied for fast inner product search~\cite{shrivastava2014asymmetric,yu2017greedy,yan2018norm,tan2019}. 

\subsubsection{Vector Compression}
Storing a high-dimensional floating-point feature vector for each of billions of ads takes a large amount of disk space and poses even more problems if these features need to be in memory for fast ranking. A general solution is compressing floating-point feature vectors into random binary (or integer) hash codes~\cite{charikar2002similarity,Proc:Li_NIPS13,DBLP:conf/nips/0001S17}, or quantized codes~\cite{jegou2011product}. The compression process may reduce the retrieval recall to an extent but it may bring significant storage benefits. For the current implementation, we adopted a quantization based method like K-Means to cluster our index vectors, rather than ranking all ad vectors in the index. When a query comes, we first find the cluster that the query vector is assigned to and fetch the ads that belong to the same cluster from the index. The idea of product quantization (PQ)~\cite{jegou2011product} goes one more step further to split vectors into several subvectors and to cluster each split separately. In our CTR model, as mentioned in Section~\ref{sec:3.1}, we split both query embeddings and ad embeddings into three subvectors. Then each vector can be assigned to a triplet of cluster centroids. For example, if we choose $10^3$ centroids for each group of subvectors, 
$10^9$ possible cluster centroids can be exploited which is adequate for a billion-scale multi-index~\cite{Lempitsky:2012:IM:2354409.2355036} for ads. In Mobius-V1, we employ a variant algorithm called Optimized Product Quantization (OPQ)~\cite{ge2013optimized}.

\begin{table*}
    \begin{center}
        \caption{Comparison results of the offline evaluation of our neural click model trained by different data generation strategies. AUC stands for the Area under the Receiver Operating Characteristic. REL is the average score of relevance between query-ad pairs in the test set, and it can be automatically evaluated by the original matching model or graded by human experts. }   
        \vspace{-0.1in}
        \begin{tabular}{l|ccc}
            \toprule
            \textbf{Neural Click Model for CTR Prediction} & \textbf{AUC} & \textbf{REL (Relevance Model)} & \textbf{REL (Human Experts)} \\
            \hline
            \hline
            2-Class (click \& unclicked Data) & \textbf{0.823} & 0.312 & 1.500\\
            3-Class (click \& unclicked Data $+$ randomly labeled bad cases) & 0.795 & 0.467 & 1.750\\ 
            3-Class (click \& unclicked Data $+$ actively learned bad cases) & 0.811 & \textbf{0.575} & \textbf{3.000}\\ 
            \bottomrule
            
        \end{tabular}
            \label{tab:eval_click}
    \end{center}
\end{table*}

\begin{table*}[!htp]
    \begin{center}
        \caption{The comparison results of different ad retrieval strategies. The experiment on ad coverage rate is conducted offline. The average response time and memory usage are tested online.}  
                \vspace{-0.1in}
        \begin{tabular}{l|cccc}
            \toprule
            \textbf{Ad Retrieval} & \textbf{Ad Coverage Rate} & \textbf{Avg. Response Time} & \textbf{Avg. Response Time} & \textbf{Memory Usage}\\
            \hline
            \hline
            Brute Force & 100\% & - & - & -\\
            Original Vector+ANN+Re-Rank &7.3\% & 120ms & 74ms & 100\%\\
            Compressed Code+MIPS  & \textbf{40.5\%} & \textbf{30ms} & \textbf{16ms} & \textbf{5\%}\\ 
            \bottomrule
            
        \end{tabular}
                        \vspace{-3mm}
            \label{tab:eval_ad}
    \end{center}
\end{table*}

\section{Experiments}
We conducted thorough experiments on Mobius-V1 before integrating it into Baidu's sponsored search engine. Specifically, we first need to conduct the offline evaluation of our CTR prediction model and the new approach on ad indexing. We need to make sure that our CTR model with the updated method on retrieving ads can discover more relevant ads with higher CPM. Then we tried to deploy it online to process a proportion of the query flow in Baidu Search. After Mobius-V1 had passed both the offline evaluation and the online A/B test, we launched it on multiple platforms to monitor the statistics of CPM, CTR and ACP (i.e., average click price).

\subsection{Offline Evaluation}

We loaded the search logs to collect the click/unclick history and built a training set which contains 80 billion samples. We also used the search log to construct the test set which has 10 billion records on ad click/unclick history.  We compare the effectiveness of our actively learned CTR model with two baseline approaches. One method is the 2-class CTR model adopted by the original ranking layer which was trained solely by the click history without using any augmented data. The other approach is a 3-class CTR model trained by the randomly augmented data without being judged by the relevance model (teacher). As shown by Table~\ref{tab:eval_click}, our model can maintain a comparable AUC with the original ranking model but significantly improves the relevance model score (from 0.312 to 0.575)  measured by our relevance model. In other words, the low-relevance but high-CPM query-ad pairs are successfully recognized as the bad cases by our new CTR model in Mobius-V1.

Moreover, we delivered the top 100,000 query-ad pairs with the highest CTR predicted by each approach to the Crowdsourcing Team in Baidu, so as to manually grade the query-ad relevance ranging from 0 to 4 (0: no relevance, 4: quite relevant) by human experts. The report of subjective opinions also demonstrates that our CTR model in Mobius-V1 performs well on discovering relevant query-ad pairs. In addition, we used the same set of to retrieve ads from two ad indexing system powered by random partition trees (ANN+Re-Rank) and OPQ (Compressed Code+MIPS), respectively. Table~\ref{tab:eval_ad} shows that OPQ increases the ad coverage rate by {\bf 33.2\%}.

\subsection{Online A/B Testing}
The online A/B testing was conducted between two different ad retrieval strategies employed by Mobius-V1 from the perspectives of average response time and memory usage. Table~\ref{tab:eval_ad} shows that OPQ can provide much lower latency than random partition tree methods and reduce the average response time by \textbf{48ms}/query. Furthermore, we examined the average response time of the top 3\% high-CPM ads which have greater business value but require more computing resources. It shows that OPQ cuts down the query latency by \textbf{75\%} (from 120ms to 30ms) and substantially saves memory consumption.  

\subsection{System Launching}
After Mobius-V1 had successfully passed both the offline evaluation and the online A/B test, we decided to launch it on multiple platforms in and outside Baidu. These platforms include the Baidu App on mobile phones, Baidu Search on PCs, and many other affiliated websites/apps that our sponsored search engine serves. Table~\ref{tab:online_eval} shows the statistics on CPM, CTR, and ACP according to our 7-day monitor on the entire online traffic. CPM is the primary metric to evaluate the performance of a sponsored search engine. Compared with the previous system, Mobius-V1 leads to a major improvement of CPM by \textbf{3.8\%} on the Baidu App and \textbf{3.5\%} on the Baidu Search, which are the main portals of our sponsored search engine.  
\begin{table}[h!]
    \begin{center}
    \caption{The improvements on CPM, CTR, and ACP of Mobius-V1 compared with the previous system deployed on different websites/apps. The results are based on our 7-day surveillance of the entire online traffic.}
                \vspace{-0.1in}
        \begin{tabular}{l|ccc}
            \toprule
            \textbf{Launched Platform} & \textbf{CPM} & \textbf{CTR} & \textbf{ACP} \\
            \hline
            \hline
            Baidu App on Mobiles & $+\textbf{3.8\%}$ & $+0.7\%$ & $+\textbf{2.9\%}$\\
            Baidu Search on PCs & $+3.5\%$ & $\textbf{+1.0\%}$ & $+2.2\%$\\
            Affiliated Websites/Apps  & $+0.8\%$ & $+0.5\%$ & $+0.2\%$\\ 
            \bottomrule
        \end{tabular}
            \label{tab:online_eval}
    \end{center}
\end{table}

\section{Related Work}
Our work on Mobius, which is towards the next-generation query-ad matching system in Baidu's sponsored search engine for commercial use, involves the research on query-ad matching and click-through rate (CTR) prediction. 

\subsection{Query-Ad Matching}
Query-ad matching \cite{raghavan2008evaluating} is an extensively studied task which aims to retrieve advertisements that are not only the same with but also semantically similar to the given query (e.g., the query ``U.S. tourist visa'' and the ads about ``travel agencies'' displayed in Figure~\ref{fig:baidu_sponsored_search}). As queries are commonly short texts, this issue has been mostly addressed by the techniques of query expansion \cite{wang2009efficient,broder2009online}, query rewriting \cite{zhang2007query,grbovic2015context} and semantic matching \cite{grbovic2016scalable,DBLP:conf/kdd/BaiOZFRST18}. Besides that we can leverage different NLP tools to directly compute the similarity between queries and textual ads, the semantic relationship between queries and ads can also be captured via learning from ad impressions. DSSM \cite{shen2014learning} is a well-known learning-to-match paradigm which leverages a deep neural architecture to capture query intent and to improve the quality of the learned semantic match given the click information.

\subsection{CTR Prediction}

CTR prediction \cite{mcmahan2013ad,tuladhar2014click} is another core task in sponsored search, as it directly influences some business indicators such as CPM. It focuses on predicting the probability that an ad would be clicked if shown as a response to a submitted query. Conventional approaches on CTR prediction preferred handcrafted features of ad impressions obtained from historical click data by Bayesian \cite{richardson2007predicting} or feature selection methods \cite{he2014practical,jahrer2012ensemble}. Along with the recent emergence of Deep Learning \cite{lecun2015deep}, many approaches \cite{zhang2014sequential,zhou2018deep,chan2018convolutional} for CTR prediction utilize various deep neural nets to primarily alleviate issues of creating and maintaining handcrafted features by learning them automatically from the raw queries and textual ads. Baidu Search Ads (``Phoenix Nest'') has been successfully using ultra-high-dimensional and ultra-large-scale deep neural networks for training CTR models since 2013.

\section{Conclusions}
In this paper, we introduce the Mobius project, which is the next generation of the query-ad matching system in Baidu's sponsored search engine, to you by answering the subsequent four questions:
\begin{itemize}
    \item \textbf{Q:} \textit{Motivation} --- why do we propose the Mobius project? \\
    
    \textbf{A}: We used to adopt a three-layer funnel-shaped structure to screen and sort hundreds of ads for display from billions of ad candidates. However, the separation/distinction between matching and ranking objectives leads to a lower commercial return. To address this issue, we set up Mobius-V1 which is our first attempt to make the matching layer take business impact measures (such as CPM) into account instead of simply predicting CTR for billions of query-ad pairs.
    \vspace{1.5mm}
    
    \item \textbf{Q:} \textit{Challenges} --- what challenges have we encountered while building Mobius-V1? \\
    
    \textbf{A:} The first problem is the insufficient click history for training the neural click model which is expected to have the generalization ability on the long-tail queries and ads. As the original neural click model employed by the ranking layer was trained by high-frequency ads and queries, it tends to estimate a query-ad pair at a higher CTR once either a high-frequency ad or a high-frequency query appears, even though they have no relevance at all. Another problem is the low retrieval efficiency and high memory consumption due to the increasing number of queries and ad candidates that Mobius has to handle. 
    \vspace{1.5mm}
    
    \item \textbf{Q:} \textit{Solutions} --- how do we design and implement Mobius to address those challenges? \\
    
    \textbf{A:} To overcome the issue of insufficiency of click history, we design a ``teacher-student'' framework inspired by active learning to augment the training data. Specifically, an offline data generator is responsible for constructing synthetic query-ad pairs given billions of user queries and ad candidates. These query-ad pairs are constantly fed into the teacher agent which is derived from the original matching layer and is good at measuring the semantic relevance of a query-ad pair. The teacher agent can help detect the bad cases (i.e., with higher CTR but lower relevance) as the augmented data from the generated query-ad pairs. Our neural click model in Mobius-V1, as a student, is taught by the additional bad cases to improve the ability of generalization. To save the computing resources and satisfy the requirement of low response latency, we tested a variety of space partitioning algorithm for the approximate nearest neighbor (ANN) search and we have found that for our datasets, OPQ~\cite{ge2013optimized} is able to achieve good performance for indexing and retrieving billions of ads more efficiently. 
    \vspace{1.5mm}
    
    \item \textbf{Q:} \textit{Feedbacks} --- how does Mobius-V1 perform in Baidu's sponsored search engine? \\
    
    \textbf{A:} We have already deployed Mobius-V1 in Baidu's sponsored search engine The results from both online and offline experiments demonstrate that this new matching system increases CPM by \textbf{3.8\%} and promotes ad coverage by {\bf 33.2\%}.
\end{itemize}

\section{Future Work}

Since 2013, Baidu Search Ads (a.k.a. Phoenix Nest) has successfully deployed ultra-large-scale deep neural networks for training CTR models. To move beyond the CTR model, Mobius has been established as an innovative and forward-looking project. The idea of unifying the objectives of optimizing the user experience and business target also inspires other featured products such as Feeds. 

For future work, many potential directions can be explored. For example,  we expect to be able to bring more business targets such as ROI (return on investment), as additional learning objectives into the matching layer so that we can discover more business-friendly ads. Along with more optimization objectives for billions of candidate ads and queries, the computational complexity will significantly increase. There ought to be a trade-off between the effectiveness and efficiency of our sponsored search engine given the requirement of lower response latency and the restraints of computing resources.

The crucial step in Mobius project is the fast ads retrieval task via approximate near neighbor search (ANN). The current system has used the cosine similarity to approximate the CTR, based on their monotonic correlation. If the final layer is more complicated, it will be problematic to rank by cosine (or weighted cosine). Searching by complicated measures has been studied, for example~\cite{tan2019}, which could be adopted by future versions of Mobius. 
Another promising direction is to adopt a GPU-based system for fast ANN, which has been shown highly effective for generic ANN tasks~\cite{Proc:Li_WWW12,JDH17,zhao2019}.

\begin{acks}
We are deeply grateful to the contributions of many colleagues from Baidu. A few names are  Lin Liu, Yue Wang, Anlong Qi, Lian Zhao, Shaopeng Chen, Hanju Guan, and Shulong Tan; but there are certainly many more who have contributed to this large project. 

\end{acks}


\balance
\bibliographystyle{ACM-Reference-Format}
\bibliography{standard}

\end{document}